\documentclass[aps,prd,twocolumn,nofootinbib,longbibliography,10pt]{revtex4-2}
\usepackage{etoolbox}
\usepackage{dcolumn}
\usepackage{amsmath,amssymb,amsfonts,mathtools,bm}
\usepackage{mathrsfs}
\usepackage{graphicx}
\usepackage[colorlinks=true,urlcolor=blue,citecolor=red,linkcolor=blue]{hyperref}
\usepackage{natbib}
\usepackage{bbold}
\usepackage{wrapfig}
\usepackage{flushend}

\makeatletter
\newsavebox{\@brx}
\newcommand{\llangle}[1][]{\savebox{\@brx}{\(\m@th{#1\langle}\)}%
  \mathopen{\copy\@brx\kern-0.5\wd\@brx\usebox{\@brx}}}
\newcommand{\rrangle}[1][]{\savebox{\@brx}{\(\m@th{#1\rangle}\)}%
  \mathclose{\copy\@brx\kern-0.5\wd\@brx\usebox{\@brx}}}
\makeatother
\begin{document}
\title{Quantum gravity signatures in gravitational wave detectors placed inside a harmonic trap potential}
\author{Soham Sen}
\email{sensohomhary@gmail.com}
\affiliation{Department of Astrophysics and High Energy Physics, S. N. Bose National Centre for Basic Sciences, JD Block, Sector-III, Salt Lake City, Kolkata-700 106, India}
\author{Sunandan Gangopadhyay}
\email{sunandan.gangopadhyay@gmail.com}
\affiliation{Department of Astrophysics and High Energy Physics, S. N. Bose National Centre for Basic Sciences, JD Block, Sector-III, Salt Lake City, Kolkata-700 106, India}
\author{Sukanta Bhattacharyya}
\email{sukanta706@gmail.com}
\affiliation{Department of Physics, West Bengal State University, Barasat, Kolkata-700 126, India}
\begin{abstract}
\noindent In this work, we consider a general gravitational wave detector of gravitational waves interacting with an incoming gravitational wave carrying plus polarization only placed inside a harmonic trap. This model can be well acquainted with the description of a resonant detector of gravitational waves as well. The well-known detector-gravitational wave interaction scenario uses the method of a semi-classical approach where the detector is treated quantum mechanically but the gravitational wave is considered at a classical level. In our analysis, we use a discrete mode decomposition of the gravitational wave perturbation which results in a Hamiltonian involving the position and momentum operators corresponding to the gravitational wave and the harmonic oscillator. We have then calculated the transition probability for the harmonic oscillator-gravitational wave tensor product state for going from an initial state to some unknown final state. Using the energy flux relation of the gravitational waves, we observe that if we consider the total energy as a combination of the number of gravitons in the initial state of the detector then the transition probability for the resonant absorption case scenario takes the analytical form which is exactly similar to the semi-classical absorption case. In the case of the emission scenario, we observe a spontaneous emission of a single graviton which was completely absent in the semi-classical analogue of this model. This therefore gives a direct signature of linearized quantum gravity.
\end{abstract}
\maketitle
\section{Introduction}
\noindent With the detection of gravitational waves by LIGO and the LIGO-VIRGO collaborations, there has been a sudden inflation in the area of research related to the quanta of the linearized gravity and its detection in the future generation of space-based gravitational wave detectors. The physics related to the graviton and detector interaction has been investigated quite thoroughly in \cite{QGravNoise,QGravLett,QGravD,Soda,
Marletto,Bose,Marletto2,Bose2,
OTMGraviton,OTMApple}. These analyses reveal the fact that a quantum gravitational treatment modifies the geodesic deviation equation by a Langevin-like equation which involves a stochastic noise term. In these works the fluctuation over the Minkowski spacetime is decomposed into individual modes of different frequencies and the modified Hamiltonian is constructed which is later raised to operator status for a quantum gravity treatment. In \cite{QGravNoise,QGravLett,QGravD,
Soda,OTMGraviton,OTMApple} the detector used is an interferometric detector that is modeled by two freely falling masses separated by a distance that is equal to the arm length of the interferometric detector. Hence, the immediate step is to consider harmonic oscillators instead of a free particle. When the gravitational wave interacts with the matter system, it creates tiny vibrations that are quite small compared to the nucleus of the atom. Such a modification resembles another kind of gravitational wave detector, also known as the Weber bar detector \cite{Weber} named after the pioneer of the ``\textit{gold rush}" of gravitational wave detection, James Weber.  Instead of Weber bar detectors, one can also consider gravitational wave interferometer detectors placed inside a harmonic trap potential.

\noindent There is a plethora of literature
\cite{Speliotopoulos,sg11,sg22,sg33,sg44,sg55,Fischer} that have investigated the quantum mechanical response of the bar detector with an incoming gravitational wave. Recently there have been some important investigations regarding parameter estimation using quantum metrological techniques \cite{Ivette1,Bruschi}. Making use of such quantum enhancement techniques, proposals of a Bose-Einstein condensate-based detection of gravitational wave as well as dark energy have been put forward in \cite{Ivette2,Ivette3,Ivette4,PhononBEC3}. All of the above investigations involve the interaction of a classical gravitational wave with quantum matter. In the analyses \cite{Speliotopoulos,sg11,sg22,sg33,sg44,sg55,Fischer} ,  for the energy scale relevant for the bar detector treated in a quantum mechanical way, the gravitational wave behaves classically. The next logical step is to consider both the harmonic oscillator as well as the gravitational wave quantum mechanically. To do so one needs to treat the small fluctuations over the background Minkowski spacetime quantum mechanically and investigate the hidden quantum gravity signatures if any for such a forced harmonic oscillator-graviton interaction model. In the semi-classical picture (where the gravitational wave is treated classically)\cite{sg22,sg33,sg44,sg55}, the transition probability for the harmonic oscillator to go from its ground state to some other excited state has been calculated using the Fermi-Golden rule. In such a case, the excitation and the absorption probabilities come out to be the same. Another important phenomenon in the scenario of the interaction of electromagnetic radiation with matter (the matter is surrounded by an electromagnetic field) is the spontaneous emission of photons. In the case of spontaneous emission of photons, a quantum mechanical system emits a quanta of energy in the form of photons and comes down from a higher excited energy level to a lower excited level. The spontaneous emission is completely a quantum field theoretic phenomenon and cannot be explained classically. We aim to investigate if such spontaneous emission can be obtained in our current analysis of the bar detector-gravitational wave system. Very recently, proposals regarding detecting single gravitons using quantum sensing have been made in \cite{TobarManikandanBeitelPikovski}.

\noindent In this work, we have considered a resonant bar detector of gravitational wave (this same model also imitates a general gravitational wave interferometer detector placed inside a harmonic trap) interacting with a quantized gravitational wave. It is important to note that we have treated both the harmonic oscillator as well as the gravitational wave quantum mechanically. Assuming the coupling constant between the gravitational wave and the detector to be small, we have used the full Hamiltonian of the system up to linear order in the coupling constant. Raising the canonically conjugate position and momentum variables to operator status for both the gravitational field and the harmonic oscillator, we obtain the final form of the Hamiltonian operator. It is now possible to separate the Hamiltonian operator into two parts, namely,  the free part of the Hamiltonian and the interaction part. Using the analytical form of the interaction part of the Hamiltonian operator in the interaction picture, we calculate the transition probability of the graviton-harmonic oscillator system for going from an initial tensor product of the individual number states to some final tensor product of similar number states. We have explicitly calculated the case where the harmonic oscillator is either initially in the ground state or the second excited state. For the resonant absorption scenario where the harmonic oscillator goes from ground state to its second excited state, one graviton gets absorbed and the gravitational wave state containing $\eta_G$ number of gravitons has now $\eta_G-1$ number of gravitons. In the $\eta_G\rightarrow 0$ limit, this transition probability vanishes which implies that when there are no gravitons in the initial state of the gravitational wave, the probability of the harmonic oscillator going from the ground state to an excited state vanishes. In the case of the harmonic oscillator being in the second excited state, we observe something very unique. In the $\eta_G\rightarrow0$ limit, the transition probability is nonvanishing and the contribution comes from the emission of a single graviton. From the equation of energy carried by gravitons, it is easy to estimate that the amplitude of the gravitational wave vanishes if the number of gravitons in a state goes to zero. Hence, for such a de-excitation, the semi-classical treatment predicts a vanishing transition probability in the $\eta_G\rightarrow 0$ limit. This implies that the harmonic oscillator can decay to its ground state by spontaneous emission of a single graviton if and only if the gravitational wave is treated as a quantum field. A very important perspective of spontaneous emission of gravitons by atomic hydrogen has been discussed in \cite{BoughnRothman}. Such spontaneous emission of gravitons has also been theoretically obtained in \cite{Pignol,OnigaWang}.

\noindent For the final part of our calculation, we have considered the energy-flux relation for gravitational waves and obtained the analytical form of the total energy of the gravitational wave in terms of the frequency and amplitude of the wave, Newton's gravitational constant and the radius of the sphere through which the energy is either received or released from the detector. We now consider that this entire energy is a combination of integral multiples of the energy carried by each quanta of the gravitational wave also known as gravitons. Hence, it is safe to substitute $E=\eta_G\hbar\omega$. Finally, we obtain an analytical expression for the wave amplitude and substitute it back in the expression for the transition probability of the resonant absorption case. Remarkably, the transition probability for the resonant absorption case matches exactly with the semi-classical analysis carried out in \cite{sg22,sg33,sg44,sg55}. This analysis implements a direct connection between the semi-classical and the quantum gravitational analysis.

\noindent 
\section{Background model and transition probability}
\noindent The background metric in which the analysis will be carried out can be thought of as a small perturbation over the Minkowski background,
\begin{equation}\label{2.1}
g_{\mu\nu}=\eta_{\mu\nu}+h_{\mu\nu}
\end{equation}
where $\eta_{\mu\nu}=\text{diag}\{-1,1,1,1\}$. A much more rigorous analysis will involve the substitution of the Minkowski metric by a post-Newtonian metric as has been done in \cite{AppleParikh}.

\noindent If we now consider the speed of light to be unity, then the Einstein-Hilbert action can be written as
\begin{equation}\label{1.2}
S_{EH}=\frac{1}{16\pi G}\int d^4 x \sqrt{-g} R
\end{equation}
with $R$ being the Ricci scalar and $g=\text{det}(g_{\mu\nu})$. Up to quadratic order in the perturbation term in Eq.(\ref{2.1}), we can recast the Einstein Hilbert action as follows
\begin{equation}\label{1.3}
\begin{split}
S_{EH}\simeq&\frac{1}{64\pi G}\int d^4x ~(h_{\mu\nu}\Box h^{\mu\nu}-h\Box h+2 h^{\mu\nu}\partial_\mu\partial_\nu h\\&-2h_{\mu\alpha}\partial_\kappa\partial^{\alpha}h^{\mu\kappa})~.\end{split}
\end{equation}
Now we shall make use of the gauge symmetry of the perturbation term given by
\begin{equation}\label{1.4}
h_{\mu\nu}=\bar{h}_{\mu\nu}+\partial_\mu\xi_\nu+\partial_\nu\xi_\mu~. 
\end{equation}
We shall now impose the transverse-traceless gauge conditions given by
\begin{align}\label{1.5}
\partial_{\kappa}\bar{h}^{\kappa\zeta}=0,~\bar{h}^{\kappa}_\kappa=0,~k_\rho\bar{h}^{\rho\zeta}=0
\end{align}
with $k_\rho=\delta_\rho^{0}$ being a constant time-like vector. 

\noindent In the transverse traceless gauge, the form of the Einstein Hilbert action in Eq.(\ref{1.3}) can be recast as
\begin{equation}\label{1.6}
S_{EH}=-\frac{1}{64\pi G}\int d^4 x~ \partial_{\kappa}\bar{h}_{ij}\partial^{\kappa}\bar{h}^{ij}.
\end{equation}
Instead of directly considering the resonant detector system, we start by considering a system of two freely falling particles with one having a much higher mass than the other particle \cite{QGravNoise,QGravLett,QGravD}. One can consider the particle with the higher mass to be on-shell. In order to proceed further, one can write down the metric in Fermi normal coordinates as follows \cite{QGravLett}
\begin{align}
g_{00}(t,\xi)&=-1-R_{i0j0}(t,0)\xi^{i}\xi^{j}+\mathcal{O}(\xi^3)~,\label{1.7}\\
g_{0k}(t,\xi)&=-\frac{2}{3}R_{0jkl}(t,0)\xi^{j}\xi^{l}+\mathcal{O}(\xi^3)~,\label{1.8}\\
g_{jk}(t,\xi)&=\delta_{jk}-\frac{1}{3}R_{jlkp}(t,0)\xi^{l}\xi^{p}+\mathcal{O}(\xi^3)\label{1.9}
\end{align}
with $\xi$ denoting the coordinate separation between the two particles.  The Riemannian tensor is evaluated on a time-like geodesic and as a result, it only depends on the temporal coordinate. Under a gauge transformation, the Riemannian tensor remains invariant for small linear perturbations about the flat Minkowski background. As a result of such small perturbations, the Riemannian tensor constructed in the Fermi-normal coordinates is the same as it is in the transverse-traceless gauge. For a detailed pedagogical derivation of the Fermi normal coordinates, we refer the readers to \cite{Ito_Soda}. 

\noindent The relativistic action for the particle with the smaller mass $m_0$ reads
\begin{equation}\label{1.10}
S_p=-m_0\int d\tau\sqrt{-g_{\mu\nu}\dot{Y}^{\mu}\dot{Y}^{\nu}}
\end{equation}
where the coordinates for the particle  are denoted by  $Y^\mu=\{t,\xi^i\}$.
One can now replace $\tau$ by $t$ in Eq.(\ref{1.10}) using the re-parametrization invariance of the action in Eq.(\ref{1.10}). The model in general represents the arm of an interferometer detector. The above action can be augmented by a harmonic potential term by placing the entire system inside a harmonic trap which represents an ideal resonating bar. The modified action reads
\begin{equation}\label{1.10a}
S_{RD}=-m_0\int dt \left(\sqrt{-g_{\mu\nu}\frac{dY^\mu}{dt}\frac{dY^\nu}{dt}}+\frac{1}{2}\omega_0^2g_{\mu\nu}Y^{\mu}Y^{\nu}\right)~.
\end{equation} 
It is important to note that the gravitational wave-particle (detector) interaction is for a very small time scale (of the order of a few milliseconds). As a result, one can also neglect terms $\mathcal{O}(t^3,t^2\xi^2)$.
Substituting Eq.(s)(\ref{1.7})-(\ref{1.9}) in Eq.(\ref{1.10a}) and keeping terms up to $\mathcal{O}(\xi^2)$, one can obtain the following simplified form of the action
\begin{widetext}
\begin{equation}\label{1.11}
\begin{split}
S_{RD}\simeq&-m_0\int dt\left(1+R_{j0k0}(t,0)\xi^j\xi^k-\delta_{ij}\dot{\xi}^i\dot{\xi}^j\right)^{\frac{1}{2}}-\frac{m_0\omega_0^2}{2}\int dt\left((1+R_{j0k0}(t,0)\xi^j\xi^k)t^2+\delta_{ij}\xi^i\xi^j\right)
\\\simeq&-m_0\int dt \Bigr(1+\frac{1}{2}R_{j0k0}(t,0)\xi^j\xi^k-\frac{1}{2}\delta_{jk}\dot{\xi}^j\dot{\xi}^k+\frac{\omega_0^2}{2}\delta_{jk}\xi^j\xi^k\Bigr)
\end{split}
\end{equation}
\end{widetext}
where $R_{j0k0}(t,0)=-\frac{1}{2}\ddot{\bar{h}}_{jk}(t,0)$ is in the transverse-traceless gauge, $\omega_0$ denotes the frequency of the harmonic trap or the bar detector, and the small terms along with the terms that will not contribute in the overall dynamics of the resonant bar have been dropped. It is important to note that the above action describing the interaction of the detector with the gravitational wave remains unchanged in the transverse-traceless gauge. 

\noindent One can now substitute the form of the Riemann curvature tensor in terms of the spacetime fluctuation ($h_{jk}$) in Eq.(\ref{1.11}) as
\begin{equation}\label{1.12}
\begin{split}
S_{RD}\simeq \frac{m_0}{2}\int dt \left[\delta_{jk}\dot{\xi}^j\dot{\xi}^k+\frac{1}{2}\ddot{\bar{h}}_{jk}(t,0)\xi^j\xi^k-\omega_0^2\delta_{jk}\xi^j\xi^k\right]
\end{split}
\end{equation}
where we have got rid of the first term in Eq.(\ref{1.11}) which will not have any contribution to the dynamics of the system. 

\noindent Our primary aim is to quantize the small spacetime fluctuations over the background spacetime metric. In order to quantize the gravitational fluctuation, we decompose the perturbation over the flat spacetime into discrete individual frequency modes, and for a valid normalization condition, we consider the entire system to be located inside a box of side length $L$. The discrete mode decomposition reads
\begin{equation}\label{2.2}
\bar{h}_{jl}(t,\bm{x})=\frac{1}{l_p}\sum\limits_{\bm{k},s}q_{\bm{k},s}e^{i\bm{k}.\bm{x}}\varepsilon^{s}_{jl}(\bm{k})
\end{equation}
where $q_{\bm{k},s}$ is the mode amplitude, $\varepsilon^{s}_{jl}$ denotes the polarization tensor with $s=+,\times$, and $\bm{k}=\frac{2\pi\bm{n}}{L}$ ($n\in \mathbb{Z}^3$) denotes the wave vector for the system being in a box of length $L$. Making use of the discrete mode decomposition in Eq.(\ref{2.2}), one can write down the full gauge fixed action of the combined gravitational wave-resonant detector system as well as interferometer detectors placed inside a harmonic trap potential as
\begin{widetext}
\begin{equation}\label{2.3}
\begin{split}
S=&S_{EH}+S_{RD}\\=&\int dt\frac{m}{2}\sum\limits_{\bm{k},s}\left(\dot{q}_{\bm{k},s}^2-\bm{k}^2q_{\bm{k},s}\right)+\int dt\frac{m_0}{2}\biggr[\delta_{jl}\dot{\xi}^j\dot{\xi}^l-\frac{1}{\sqrt{\hbar G}}\sum\limits_{\bm{k},s}\dot{q}_{\bm{k},s}\varepsilon^{s}_{lj}(\bm{k})\dot{\xi}^{j}\xi^l-\omega_0^2\delta_{jl}\xi^j\xi^l\biggr]
\end{split}
\end{equation}
\end{widetext}
where $\xi$ is the coordinate variables corresponding to the detector phase space, $\omega_0$ is the frequency of the detector (harmonic oscillator frequency), and $m=\frac{L^3}{16\pi\hbar G^2}$. For a resonant bar detector, a one-dimensional model can be considered, as the length of the bar is always much larger than the other two directions perpendicular to its length. For an interferometer detector, the long detector arm can be considered to be one-dimensional (a two-mass system with one mass way smaller than the other one) and the harmonic trap potential term will be taken care of by its one-dimensional analogue. It is now possible to consider the wave vector $\bm{k}$ to be propagating along the $z$ direction with magnitude $\omega=|\bm{k}|$, along with plus polarization only. Considering negligible fluctuations along the $z$ and $y$ directions, one can recast the total action of the system as
\begin{equation}\label{2.4}
S=\int dt\left[\frac{m}{2}\left(\dot{q}^2-\omega^2q^2\right)+\frac{m_0}{2}\left(\dot{\xi}^2-\frac{2\mathcal{G}\dot{q}\dot{\xi}\xi}{m_0}-\omega_0^2\xi^2\right)\right]
\end{equation} 
where $\mathcal{G}=\frac{m_0}{2\sqrt{\hbar}G}$, $\xi^{x}=\xi$ and $\Re(q_{k_z},+)=q$. It is important to note that we have considered $c=1$ in our current analysis. The Lagrangian corresponding to the action in the above equation reads
\begin{equation}\label{1.16}
L=\frac{m}{2}\left(\dot{q}^2-\omega^2q^2\right)+\frac{m_0}{2}\left(\dot{\xi}^2-\omega_0^2\xi^2\right)-\mathcal{G}\dot{q}\dot{\xi}\xi~.
\end{equation}
The momentum conjugate to the position variables $q$ and $\xi$ read
\begin{align}
p&=\frac{\partial L}{\partial \dot{q}}=m\dot{q}-\mathcal{G}\dot{\xi}\xi~,~
\pi&=\frac{\partial L}{\partial\dot{\xi}}=m_0\dot{\xi}-\mathcal{G}\dot{q}\xi~.\label{1.17}
\end{align}
Using the form of the Lagrangian in Eq.(\ref{1.16}) and the conjugate momentum variables, one can write down the Hamiltonian for the system as
\begin{equation}\label{1.18}
H=\frac{\frac{p^2}{2m}+\frac{\pi^2}{2m_0}+\frac{\mathcal{G}p\pi\xi}{mm_0}}{1-\frac{\mathcal{G}^2\xi^2}{mm_0}}+\frac{1}{2}m\omega^2q^2+\frac{1}{2}m_0\omega_0^2\xi^2~.
\end{equation}
As the coupling constant $\mathcal{G}$ is assumed to be very small, one can express the Hamiltonian of the system in Eq.(\ref{1.18}) up to $\mathcal{O}\left(\mathcal{G}\right)$ as
\begin{equation}\label{1.19}
H\simeq\frac{p^2}{2m}+\frac{\pi^2}{2m_0}+\frac{\mathcal{G}p\pi\xi}{mm_0}+\frac{1}{2}m\omega^2q^2+\frac{1}{2}m_0\omega_0^2\xi^2~.
\end{equation}
Following \cite{QGravD}, we also assume that the coupling $\mathcal{G}$ is turned on and off adiabatically and as a result $\mathcal{G}(t)\rightarrow\mathcal{G}f(t)$ such that $f(t<t_i)=f(t>t_f)=0$ and $f(t_i\leq t\leq t_f)=1$. Here, $t_i$ denotes the time at which the interaction between the detector and the gravitational wave begins and $t_f$ denotes the time when the gravitational wave stops interacting with the detector. This switching function allows one to define tensor product states at times $t=t_i$ and $t=t_f$. 

\noindent In order to quantize the above Hamiltonian,  we raise the position and momentum variables corresponding to both the resonant detector and gravitational wave to operator status and obtain the Hamiltonian operator of the detector-graviton system to be
\begin{equation}\label{2.5}
\begin{split}
&\hat{H}=\left(\frac{\hat{p}^2}{2m}+\frac{1}{2}m\omega^2\hat{q}^2\right)\otimes\hat{\mathbb{1}}_{RD}\\&+\hat{\mathbb{1}}_{GW}\otimes\left(\frac{\hat{\pi}^2}{2m_0}+\frac{1}{2}m_0\omega_0^2\hat{\xi}^2\right)+\frac{\mathcal{G}}{2mm_0}\hat{p}\otimes\left(\hat{\xi}\hat{\pi}+\hat{\pi}\hat{\xi}\right)
\end{split}
\end{equation} 
where $[\hat{\xi},\hat{\pi}]=[\hat{q},\hat{p}]=i\hbar$ with $\hat{\mathbb{1}}_{RD}$ and $\hat{\mathbb{1}}_{GW}$ denoting the identity operators corresponding to the Hilbert space of the resonant bar detector\footnote{In order to avoid making repeated statements, we have stuck to the resonant bar detector model system as our primary model. It is also important to note that the same things will hold for an interferometer detector placed inside a harmonic trap potential.} and the gravitational wave.  The Hamiltonian in Eq.(\ref{2.5}) can be separated into two parts, $\hat{H}=\hat{H}_0+\hat{H}^{\text{int}}$ with $\hat{H}^{\text{int}}=\frac{\mathcal{G}}{2mm_0}\hat{p}\otimes\left(\hat{\xi}\hat{\pi}+\hat{\pi}\hat{\xi}\right)$ denoting the interaction part of the Hamiltonian. Before the gravitational wave has started interacting with the detector and after it has stopped interacting, we can consider the total state of the system as a tensor product of the individual number states corresponding to the gravitational wave and the harmonic oscillator \cite{QGravD}. The position and momentum operators for the gravitational wave as well as the resonant bar detector in terms of their respective creation and annihilation operators read
\begin{align}
\hat{q}&=\sqrt{\frac{\hbar}{2m\omega}}\left(\hat{a}+\hat{a}^\dagger\right)~,~\hat{p}=i\sqrt{\frac{m\hbar\omega}{2}}\left(\hat{a}^\dagger-\hat{a}\right)~;\label{2.6}\\
\hat{\xi}&=\sqrt{\frac{\hbar}{2m_0\omega_0}}\left(\hat{\chi}+\hat{\chi}^\dagger\right)~,~\hat{\pi}=i\sqrt{\frac{m_0\hbar\omega_0}{2}}\left(\hat{\chi}^\dagger-\hat{\chi}\right)\label{2.7}
\end{align}
where $[\hat{a},\hat{a}^\dagger]=1=[\hat{\chi},\hat{\chi}^\dagger]$. 

\noindent Using Eq.(s)(\ref{2.6},\ref{2.7}) and making use of the commutation relation, one can recast Eq.(\ref{2.5}) in the following form
\begin{equation}\label{2.7a}
\begin{split}
\hat{H}=&\hbar\omega\left(\hat{a}^{\dagger}\hat{a}+\frac{1}{2}\right)\otimes\hat{\mathbb{1}}_{RD}+\hat{\mathbb{1}}_{GW}\otimes\hbar\omega\left(\hat{\chi}^{\dagger}\hat{\chi}+\frac{1}{2}\right)\\
&-\frac{\hbar\mathcal{G}}{2mm_0}\sqrt{\frac{m\hbar\omega}{2}}\left(\hat{a}^{\dagger}-\hat{a}\right)\otimes\left(\hat{\chi}^{\dagger2}-\hat{\chi}^2\right)~.
\end{split}
\end{equation}

\noindent One can now define the eigenstates corresponding to the individual number operators $\hat{N}_{GW}=\hat{a}^\dagger\hat{a}$ and $\hat{N}_{RD}=\hat{\chi}^{\dagger}\hat{\chi}$ as 
\begin{equation}\label{2.8}
\hat{N}_{GW}|\eta_G\rangle=\eta_G|\eta_G\rangle~,~\hat{N}_{RD}|n_r\rangle=n_r|n_r\rangle
\end{equation}
where $|n_r\rangle$ denotes the $n_r$-th excited state of the harmonic oscillator and $|\eta_G\rangle$ denotes the gravitational wave state containing $\eta_G$ number of gravitons. As the initial and final state of the gravity waver-detector system can be represented as a tensor product of the corresponding number states, we can write the initial and final state of the system as $|\psi_i\rangle=|\eta_G\rangle\otimes|n_r\rangle=|\eta_G,n_r\rangle$ and $|\psi_f\rangle=|\eta'_G,n'_r\rangle$ with $|\psi_i\rangle\neq |\psi_f\rangle$. Our primary aim is to calculate the transition probability for the system going from the state $|\psi_i\rangle$ to $|\psi_f\rangle$. In order to calculate the form of the transition probability we need the analytical structure of the interaction Hamiltonian in the interaction picture which is calculated as
\begin{equation}\label{2.9}
\begin{split}
\hat{H}^{\text{int}}_I(t)&=e^{\frac{i}{\hbar}\hat{H}_0t}\hat{H}^{\text{int}}e^{-\frac{i}{\hbar}\hat{H}_0t}\\&=\frac{\mathcal{G}}{2mm_0}\hat{p}_I\otimes\left[\hat{\xi}_I\hat{\pi}_I+\hat{\pi}_I\hat{\xi}_I\right]
\end{split}
\end{equation}
where $\hat{p}_I(t)$, $\hat{\xi}_I(t)$, and $\hat{\pi}(t)$ are $\hat{p}$, $\hat{\xi}$, and $\hat{\pi}$ respectively in the interaction picture. 

\noindent Upto first order in the interaction Hamiltonian in the interaction picture, one can write down the transition probability of the system for going from $|\psi_i\rangle$ to $|\psi_f\rangle$ as follows
\begin{equation}\label{2.10}
\begin{split}
P_{if}(t)&=\left|\langle\psi_f|\hat{U}_I(t,t_i)|\psi_i\rangle\right|^2\\
&\simeq\frac{1}{\hbar^2}\left|\int_{t_i}^{t}dt'\langle\psi_f|\hat{H}^{\text{int}}_I(t')|\psi_i\rangle\right|^2~.
\end{split}
\end{equation}
The usual notion of calculating such transition probabilities is to set the limit $t_i\rightarrow-\infty$ and $t_f\rightarrow \infty$ \cite{sg22,sg33}. Extending the integration limits and substituting the form of $\hat{H}^{\text{int}}_I(t')$ from Eq.(\ref{2.9}) in the above equation, we obtain the form of the transition probability as
\begin{equation}\label{2.11}
\begin{split}
&P_{if}=\frac{1}{\hbar^2}\left|\int_{-\infty}^{\infty}dt'\langle\psi_f|\hat{H}^{\text{int}}_I(t')|\psi_i\rangle\right|^2\\
&=\frac{\hbar\omega\pi^2\mathcal{G}^2}{2mm_0^2}\Bigr|\sqrt{(\eta_G+1)(n_r+2)(n_r+1)}\delta_{\eta'_G,\eta_G+1}\delta_{n_r',n_r+2}\\&\times\delta(\omega+2\omega_0)-\sqrt{(\eta_G+1)n_r(n_r-1)}\delta_{\eta'_G,\eta_G+1}\delta_{n_r',n_r-2}\\&\times\delta(\omega-2\omega_0)-\sqrt{\eta_G(n_r+2)(n_r+1)}\delta_{\eta'_G,\eta_G-1}\delta_{n_r',n_r+2}\\&\times\delta(-\omega+2\omega_0)+\sqrt{\eta_Gn_r(n_r-1)}\delta_{\eta'_G,\eta_G-1}\delta_{n_r',n_r-2}\\&\times\delta(-\omega-2\omega_0)\Bigr|^2
\end{split}
\end{equation}
where $\int_{-\infty}^\infty dt e^{i(\omega_1-\omega_1')}=2\pi\delta(\omega_1-\omega_1')$. Now $\delta(\omega+2\omega_0)$ (or $\delta(-\omega-2\omega_0)$) gives non vanishing contribution when $\omega=-2\omega_0$ which is not a physical condition. Therefore, the transition probability in Eq.(\ref{2.11}) can be simplified as follows
\begin{equation}\label{2.12}
\begin{split}
&P_{if}=\frac{\hbar\omega\pi^2\mathcal{G}^2}{2mm_0^2}\Bigr(\sqrt{(\eta_G+1)n_r(n_r-1)}\delta_{\eta_G',\eta_G+1}\delta_{n_r',n_r-2}\\&+\sqrt{\eta_G(n_r+2)(n_r+1)}\delta_{\eta_G',\eta_G-1}\delta_{n_r',n_r+2}\Bigr)^2\delta^2(\omega-2\omega_0)~.
\end{split}
\end{equation}
The above form of transition probability ensures the fact that due to graviton-detector interaction, the harmonic oscillator will always jump up or jump down two consecutive energy eigenstates. 
\section{Resonant absorption}
\noindent We consider the simple case of the resonant bar detector being in a ground state ($n_r=0$). From the form of the transition probability in Eq.(\ref{2.12}), it is evident that only the second term within the parenthesis contributes towards a non-vanishing probability for the system provided that $\eta_G'=\eta_G-1$. The condition $\eta_G'=\eta_G-1$ signifies the fact that the final gravitational wave state is short of a graviton that has been absorbed by the detector and the detector has excited to the second excited energy level simultaneously. For $n_r=0$, we obtain the form of the transition probability as
\begin{equation}\label{2.13}
P_{02}=\frac{\eta_G\hbar\omega\pi^2\mathcal{G}^2}{mm_0^2}\delta^2(\omega-2\omega_0)=\frac{4\eta_G\hbar\omega \pi^3G}{L^3}\delta^2(\omega-2\omega_0)
\end{equation}   
where we have substituted the analytical forms of $\mathcal{G}$ and $m$. Our next aim is to establish a connection between the transition amplitude obtained in our case with that of the semi-classical case where the phase-space variables obey Heisenberg's uncertainty principle. In \cite{sg22,sg33}, the transition probability for the harmonic oscillator going from its ground state to a higher excited state was calculated for a generalized uncertainty principle (GUP) framework. If now the GUP parameter is taken to zero, only the transition probability indicating a ground state to second excited state transition, survives. The transition probability in such a semiclassical  analysis for a periodic linearly polarized gravitational wave with the form $h_{jk}(t)=2f_0\cos\omega t(\varepsilon_\times\sigma^{1}_{jk}+\varepsilon_+\sigma^{3}_{jk})$, takes the form
\begin{equation}\label{2.14}
P_{02}=\frac{1}{2}\pi^2f_0^2\omega^2\varepsilon_+^2\delta^2(\omega
-2\omega_0)
\end{equation}
with $f_0$ denoting the amplitude of the gravitational wave.
For the case of the gravitational wave carrying a plus polarization only, one obtains $\varepsilon_+=1$ (using the condition $\varepsilon_+^2+\varepsilon_\times^2=1$). In order to truly relate the quantum gravitational and semi-classical results, we need to calculate the energy carried by the gravitational wave.

\noindent The total energy flowing through an area $dA$ within the time span $t=-\infty$ to $t=\infty$ reads \cite{Maggiore}
\begin{equation}\label{2.15}
\frac{dE}{dA}=\frac{1}{32\pi G}\int_{-\infty}^\infty dt\langle \dot{h}_{ij}^{TT}\dot{h}_{ij}^{TT}\rangle 
\end{equation}
where $h_{ij}^{TT}$ denotes the fluctuating part of the spacetime metric in the transverse traceless gauge and $\langle\cdots\rangle$ denotes a temporal average. The above formula assumes the entire system inside a sphere of radius $r$, hence, one can write $dA=r^2d\Omega$ which indicates the energy going out or coming in through a solid angle $d\Omega$. As our entire analysis is restricted to one dimension only, the $h^{TT}_{11}$ component will contribute in Eq.(2.15), leading to the expression of the total energy leaving the sphere or vice-versa as
\begin{equation}\label{2.16}
E=\frac{r^2}{8G}\int_{-\infty}^{\infty} dt \langle\dot{h}_+^2(t)\rangle~.
\end{equation}
Now the integral can be conducted first then the temporal average is just an average over a constant parameter \cite{Maggiore}. Hence, one can get rid of the average. Now for a periodic plane polarized gravitational wave with a template $h_+(t)=2f_0\cos\omega t$, a single time cycle ranges from $t=0$ to $t=\frac{2\pi}{\omega}$. It is now possible to calculate the total energy entering or exiting through the spherical area in one periodic cycle as
\begin{equation}\label{2.17}
E=\frac{r^2f_0^2\omega^2}{2G}\int_0^{\frac{2\pi}{\omega}}\sin^2\omega t=\frac{\pi\omega r^2 f_0^2}{2G}~.
\end{equation}
 Although the entire system is effectively one-dimensional in our analysis (because of the larger length of the detector than its span in other directions), when the energy is emitted or released via the enclosed (imaginary) sphere, we need to consider a three-dimensional post-interaction picture. Initially, we have considered a box of length $L$ but when the energy flux relation is considered, a sphere of radius $r$ is considered. Effectively, for a large $L$ we can consider the box of length $L$ to be embedded inside the sphere of radius $r$ such that all the corner points of the box lie on the surface of the sphere. Under such an assumption, $r=\frac{L}{\sqrt{2}}$ and the energy expression takes the form $E=\frac{\pi\omega f_0^2L^2}{4G}$. If the gravitational wave is quantized, then the energy $E$ will be carried by $\eta_G$ number of gravitons with mode frequency $\omega$. Hence, we can write $E=\eta_G\hbar\omega$. Substituting the value of $r$ and $E$ in Eq.(\ref{2.17}), we obtain the form of the square of the amplitude as
\begin{equation}\label{2.18}
\begin{split}
f_0^2=\frac{4\eta_G\hbar G}{\pi L^2}~.
\end{split} 
\end{equation}  
Using the above form of the square of the amplitude in Eq.(\ref{2.13}), we can write down the form of the transition probability, $P_{02}$ as
\begin{equation}\label{2.19}
P_{02}=\frac{f_0^2\pi^4\omega}{L}\delta^2(\omega-2\omega_0).
\end{equation}
In order to find out the final expression for the transition probability $P_{02}$, we need the analytical expression of the box length $L$. Initially, when we have considered a box of length $L$, the underlying assumption is that the interaction and its effects are confined within the box of length $L$. Now in time $t=\frac{2\pi}{\omega}$, the gravitational wave travels a path of distance $\lambda=\frac{2\pi c}{\omega}\bigr\rvert_{c\rightarrow1}=\frac{2\pi}{\omega}$. Hence, one can consider a sphere of radius $\lambda$ around the harmonic oscillator such the oscillator is placed at the center of the sphere. Again, for a sufficiently large $L$, we have assumed a spherical enclosure so that the initial box of length $L$ is stretched over a sphere of radius $\lambda$. 
 Hence, one can make the simple assumption $L\simeq\pi \lambda=\frac{2\pi^2}{\omega}$. Substituting the value of $L$ in Eq.(\ref{2.19}), we obtain the form of the transition probability to be
\begin{equation}\label{2.20}
P_{02}=\frac{1}{2}f_0^2\pi^2\omega^2\delta^2(\omega-2\omega_0)
\end{equation}
which is identical to the form of the transition probability in Eq.(\ref{2.14}) ($\varepsilon_+=1$ as $\varepsilon_\times=0$) obtained using the semi-classical treatment. Here we have made use of the energy-flux relation of a classical gravitational wave in Eq.(\ref{2.15}) and have considered the total energy to be carried by $\eta_G$ number of gravitons. Then we used it in the quantum gravity scenario to relate the quantum gravitational results with that of the semi-classical analysis. The striking feature is the exact similarity of the two results. This exact identification with the semi-classical treatment ensures the fact that the consideration of the gravitational wave with an integral number of gravitations carrying energy equal to the reduced Planck's constant multiplied by the frequency of the gravitational wave is quite valid. Now, although the absorption case is identical to the semi-classical analogue of this model, the emission case offers a little bit more subtlety. This we shall look at in the next section.
\section{Spontaneous emission of gravitons}
\noindent We now consider the case when the initial state of the total system is $|\psi_i\rangle=|\eta_G,2\rangle$ and the final state is $|\psi_f\rangle=|\eta_G+1,0\rangle$. The form of the transition probability in Eq.(\ref{2.12}) for the above initial and final states take the form
\begin{equation}\label{2.21}
\begin{split}
P_{20}&=(\eta_G+1)\frac{4\hbar\omega G\pi^3}{L^3}\delta^2(\omega-2\omega_0)~.
\end{split}
\end{equation}
For the semi-classical scenario, it is easy to observe that $P_{02}=P_{20}$ \cite{sg11,sg22,sg33,sg44,sg55}. If $\eta_G\rightarrow 0$, then from Eq.(\ref{2.18}) it is easy to see that $f_0$ vanishes and as a result $P_{02}$ also vanishes. But in this limit $P_{20}\neq 0$, and takes the form
\begin{equation}\label{2.22}
P_{20}=\frac{h\omega^4 G}{4\pi^4}\delta^2(\omega-2\omega_0)~.
\end{equation}
Now $\eta_G\rightarrow 0$ signifies that the initial state of the gravitational wave-detector system is $|\psi_i\rangle=|0,2\rangle$ denoting that there are no gravitons in the initial tensor product state. The non-vanishing probability in Eq.(\ref{2.22}) indicates that when the harmonic oscillator is in an excited state, it can emit a graviton spontaneously and come back to the ground state. This spontaneous emission is a direct consequence of a degenerate parametric down conversion process\footnote{A parametric down-conversion process is a non-linear optical process (in this case gravitational) where two or more excitations spontaneously convert into a single photon (here graviton) or vice versa.} as can be seen from the form of the interaction term in Eq.(\ref{2.7a}) and as a result it creates an asymmetry between the resonant absorption and spontaneous emission processes. This spontaneous emission of graviton is purely a quantum gravity phenomenon and is absent in the semi-classical analysis. As the gravitational wave is treated as a quantum field, $\eta_G\rightarrow 0$ condition implies that the initial state is a vacuum state. The spontaneous emission scenario of gravitons can be thought of as vacuum fluctuation of the surrounding gravitational field (same as the electromagnetic counterpart). The general frequency range of detection by a LIGO-VIRGO detector is in the $10^2-10^4$ Hz range. In the case of the Weber bar detectors, they can reach up to a resonance frequency $\omega_0\sim 5\times10^3$ Hz. The resonant condition ensures that $\omega\sim 10^4$ Hz. For such a case, the transition probability in Eq.(\ref{2.22}) takes the form $P_{02}=\frac{h\omega^4 G}{4\pi^4 c^5}\delta^2(\omega-2\omega_0)\sim (10^{-73}\text{sec}^{-2}) \delta^2(\omega-2\omega_0)$~. For a classical gravitational wave $f_0\sim 10^{-21}$ which gives a value of the transition probability in Eq.(\ref{2.20}) as $P_{02}\sim (10^{-33}\text{sec}^{-2})\delta^2(\omega-2\omega_0)$. Equating the above result and the contribution from $\eta_G$ number of gravitons indicates that for $f_0\sim 10^{-21}$ there are $\eta_G\sim10^{40}$ number of gravitons carrying the energy $E$ in the time interval $0<t<\frac{2\pi}{\omega}$. It is important to note that every bar detector can be considered as a combination of a large number of cylinders of harmonic oscillators of length equal to the length of the resonant bar. Now the diameter of such a cylinder can attain the lowest value which is equal to the diameter of a single atom. A Weber bar now has a diameter of one meter. The radius of an atom is close to $r_0\sim 3\times10^{-10}m$. Hence, there is approximately $(\pi 0.5^2)/(\pi (3\times 10^{-10})^2)\sim 10^{18}$ number of atoms on the surface of the bar detector. Therefore, the entire bar detector can be considered as a combination of maximum $N\sim10^{18}$ numbers of such one-dimensional cylinders with harmonic oscillator frequency equal to $\omega_0$. Then each bar detector has a mass $m_0'=m_0/N$. As they are of the same length and made of the same components we can assume the frequencies of each of the oscillators to be the same. Hence, the transition probability due to spontaneous emission from one such harmonic oscillator gets multiplied by $10^{18}$ resulting in a joint transition probability of $P_{02}=(10^{-55}\text{sec}^{-2})\delta^2(\omega-2\omega_0)$. Although the existence of the square of the Dirac-delta function theoretically claims that it will make the transition probability very high when the resonance condition gets satisfied, the Dirac-delta function comes in due to a time integral from $t_i\rightarrow-\infty$ to $t\rightarrow\infty$ in the form of the transition probability in Eq.(\ref{2.10}). A more realistic scenario will occur for a finite initial and final limit of the time integral in the form of the transition probability. Now a spontaneous emission of graviton denotes that the initial vacuum state will have one graviton. If the number of spontaneous emissions is made arbitrarily high, the final field state will contain a large number of gravitons which will effectively create a very small fluctuation over the background. This can be considered as a gravitational fluorescence-like effect where gravitons will emit in all possible directions. Still, a collective spontaneous emission shall lead to such amplified fluctuation over the spacetime metric. Such fluctuations, if they exist, will be almost impossible to detect. Now, if it is possible to modify the bar detectors in the future in such a way that the resonance condition amplifies the transition probability exponentially, then it may be possible to detect such spontaneous emission from resonant bar detectors in a very far future. It may be also possible to detect this kind of scenario in an interferometer detector if it is possible to create a harmonic trap potential for the same.
\section{Conclusion}
\noindent We consider a resonant bar detector of gravitational wave (or an interferometer detector placed inside a harmonic trap potential) interacting with an incoming gravitational wave with plane polarization only. We apply a mode decomposition of the small perturbation over the Minkowski background and obtain the total Hamiltonian of the gravitational wave-resonant bar detector system. The Hamiltonian operator is then constructed by raising all the phase space variables corresponding to the gravitational wave as well as the resonant bar detector. Making use of the Fermi-Golden rule, we obtain the transition probability of the system going from an initial joint tensor product of individual number states to some final state. We then specifically consider the case for the detector going from the ground state to the second excited energy state. Using the energy flux relation for a classical gravitational wave and considering the total energy carried by the gravitational wave as a combination of $\eta_G$ number of gravitons, we observe that the transition probability is exactly similar to that of the case where a classical gravitational wave interacts with a resonant bar detector. This identification with the semi-classical model solidifies the quantum gravity approach of using the graviton-bar detector interaction model. For a semi-classical model, the resonant absorption and emission probabilities are exactly the same. However, in the current quantum gravity analysis, we observe that for a de-excitation of the resonant bar detector from the second excited energy state to the ground state, the transition probability does not vanish for $\eta_G\rightarrow 0$ limit contrary to the resonant absorption case. This indicates the phenomenon of spontaneous emission of a graviton. This is the most interesting outcome of our analysis as spontaneous emission of gravitons cannot be observed in a semi-classical let alone a classical analysis. It is important to note that some very small local perturbations can replicate such a spontaneous emission process. Hence, one needs to construct an experimental scenario such that no other external perturbations lead to a similar effect. Under such conditions, if the spontaneous emission of gravitons is detected it can be considered to be a possible evidence for a signature of quantum gravity. Finally, we have calculated some numerical values indicating that it will be very difficult to observe the spontaneous emission of gravitons from a resonant bar detector unless the resonance condition amplifies the transition probability exponentially.

\end{document}